\documentclass[aps,prl,twocolumn,superscriptaddress,floatfix,amsmath,amssymb,hyperref]{revtex4-2}

\usepackage{graphicx} 
\usepackage{amsthm}
\usepackage{dcolumn} \usepackage{bm} \usepackage{tikz} \usetikzlibrary{arrows.meta,fadings}
\theoremstyle{remark}
\newtheorem{remark}{Remark}
\begin{document}

\title{Thermodynamic Invariants of Coupled Channels: A Many-Channel Tolman--Ehrenfest Effect}

\author{Benjamin Hamblin}
  \email{Benjamin.Hamblin@Curtin.edu.au}
\author{Victor Calo}
  \email{Victor.Calo@Curtin.edu.au}
\affiliation{ARC Centre of Excellence for Carbon Science \& Innovation,
             Perth, WA~6102, Australia}
\affiliation{EECMS: Applied Mathematics,
             Curtin University, Perth, WA~6102, Australia}
\author{Klaus Regenauer-Lieb}
  \email{Klaus@Curtin.edu.au}
\affiliation{ARC Centre of Excellence for Carbon Science \& Innovation,
             Perth, WA~6102, Australia}
\affiliation{WASM: Minerals, Energy and Chemical Engineering,
             Curtin University, Perth, WA~6102, Australia}

\date{\today}

\begin{abstract}
When multiple thermodynamic channels are coupled, single-channel equilibrium conditions fail. Extending the Tolman--Ehrenfest effect to the entropy manifold, we derive the unique $n$-channel invariant $\zeta_i T_i = C$, where $\zeta_i$ is the holonomy of the Ruppeiner connection. For the granular volume--stress ensemble, Rowe's dilatancy ratio and energy restriction emerge as geometric consequences of the off-diagonal curvature $g_{V\sigma}$, and the 60-year puzzle of state-dependent $K_\mu$ is resolved: the correction $\zeta_V$ reaches $O(1)$ near jamming. The prediction $\zeta_V\chi=\mathrm{const}$ across a shear band is experimentally testable.
\end{abstract}

\maketitle

\textit{Introduction.}---Classical thermodynamics achieves its economy by assuming the entropy surface is effectively diagonal: channels equilibrate independently.  For dense granular media, reactive geophysical fluids, and active matter~\cite{Scheibner2020,Fruchart2021}, coupling between thermodynamic channels is the dominant physics. Existing frameworks include Blumenfeld--Edwards granular statistical mechanics~\cite{Blumenfeld,Powders}, gauge-theoretic thermodynamics~\cite{Thermo-gauge}, and the cross-diffusion reduction of Biktashev and Tsyganov~\cite{BiktashevTsyganov2016}. These approaches identify the need for coupling corrections but do not derive the invariant that coupled channels share at equilibrium.

The Ruppeiner metric~\cite{Ruppeiner1979,Ruppeiner1995}, $g_{ij}=-\partial^2 S/\partial q^i\partial q^j$, quantifies this failure: $g_{ij}$ ($i\neq j$) measures how much injecting $q^j$ shifts the conjugate intensity $T_i$ at fixed $q^i$, so that $\partial\beta_i/\partial q^j=-g_{ij}$.  It is the precision matrix of equilibrium fluctuations and diverges at critical points~\cite{Ruppeiner1995}.  The Ruppeiner and Weinhold~\cite{Weinhold1975} metrics have been applied extensively to phase transitions, black hole thermodynamics, and information-geometric characterisation of thermodynamic systems~\cite{BrodyRiegler2003,Crooks2007}; what has not been identified in this literature is the specific \emph{invariant} that replaces the bare intensive variable when channels are coupled. 
In the relativistic Tolman--Ehrenfest (TE) effect~\cite{Tolman1930} the conserved quantity in a gravitational field is the redshifted energy, forcing $T(1+\phi/c^2)=\mathrm{const}$; here inter-channel coupling plays the role of the gravitational potential, and the geometric weight $\zeta_i$ is computed from the entropy-surface curvature.

\textit{MCTE relation.}---Let a body $\mathcal{B}$ possess $n$ thermodynamic channels with extensive variables $\boldsymbol{q}=(q^1,\dots,q^n)$ and conjugate intensities $\beta_i=\partial S/\partial q^i$, $T_i=\beta_i^{-1}$.  The correct multi-channel equilibrium condition must be a \emph{first integral} of the level-set flow on the entropy manifold: a function of the intensities $\{\beta_i\}$ that is constant along every trajectory of the flow induced by the constraint $S=\mathrm{const}$.

On the level set $dS=\sum_i\beta_i\,dq^i=0$, the Ruppeiner metric $\partial\beta_i/\partial q^j=-g_{ij}$ drives the coupled flow
\begin{equation}
  \left.\frac{d\beta_i}{dq^j}\right|_{dS=0}
  = \frac{g_{ii}\beta_j - g_{ij}\beta_i}{\beta_i},
  \quad j\neq i.
  \label{eq:levelset-flow}
\end{equation}
A first integral $F(\beta_1,\dots,\beta_n)$ of Eq.~(\ref{eq:levelset-flow}) must satisfy $\nabla_{\!\beta}F\cdot \dot{\boldsymbol{\beta}}=0$ along every level-set trajectory for all $g_{ij}$.  The unique solution of this integrability condition, derived in SM1 (Supplemental Material) by separation of variables on the level-set ODE is $F=\zeta_i\beta_i^{-1}$ for all~$i$, where $\zeta_i$ satisfies
\begin{equation}
  d\log\zeta_i = \sum_{j\neq i}
    \frac{g_{ii}\beta_j - g_{ij}\beta_i}{\beta_i^2}\,dq^j
    =: \omega_i.
  \label{eq:omega}
\end{equation}
Setting $F=C$ then gives the MCTE relation
\begin{equation}
  \boxed{\zeta_i\,T_i = C \quad \text{on } \mathcal{B}, \quad
    i=1,\dots,n,}
  \label{eq:MCTE}
\end{equation}
where $C$ is a single scalar invariant across all channels and all points: the multi-channel generalisation of the relativistic Tolman temperature $T\sqrt{g_{00}}$.   The multiplicative effective energy $E_{\rm eff}=\int\zeta_a e_a\,dv$ whose conservation enforces Eq.~(\ref{eq:MCTE}) is not assumed but follows from the first-integral requirement (see SM1 in Supplemental Material~\cite{supplement}).  
Equation~(\ref{eq:MCTE}) is also the Euler--Lagrange equation of $\mathcal{L}[e_a]=\int s\,dv-\beta\int \zeta_a e_a\,dv$, confirming $C=\beta^{-1}$ (see SM2 in Supplemental Material~\cite{supplement}).

\textit{Coupling coefficient as Ruppeiner holonomy.}---The MCTE condition requires $d\log\zeta_i=d\log\beta_i$ along every entropy level set.  The one-form $\omega_i$ defined in Eq.~(\ref{eq:omega}) integrates to give the explicit coupling coefficient
\begin{equation}
  \zeta_i(\boldsymbol{q}) = \lambda_i\exp\!\left(
    \int_{\boldsymbol{q}_0}^{\boldsymbol{q}} \omega_i
  \right),
  \label{eq:zeta-explicit}
\end{equation}
with $\lambda_i$ fixed by a single boundary condition.  When $g_{ij}=0$ for all $i\neq j$ and $g_{ii}$ is position-independent, $\zeta_i=\lambda_i$ (constant) and Eq.~(\ref{eq:MCTE}) reduces to the classical $T_i=\mathrm{const}$.

The one-form $\omega_i$ is the projection of the logarithmic Jacobian of the intensive-variable map onto the entropy level set.  The integral $\oint\omega_i$ around a closed loop is the \emph{holonomy} of the Ruppeiner connection: it vanishes for a flat metric (decoupled channels) and is non-zero whenever inter-channel coupling is present, measuring the ``thermodynamic twist'' accumulated along a closed equilibrium path.  In the Tolman--Ehrenfest analogy, $\zeta_i$ corresponds to $\sqrt{g_{00}}$, which is the norm of the time-translation Killing vector; here it is the norm of the vector field generating translations along effective-energy level sets on the entropy manifold.

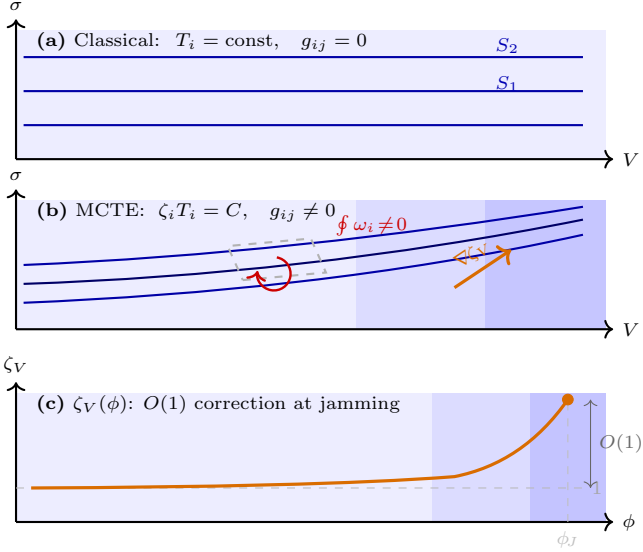
\begin{figure}[tb]
\centering
\begin{tikzpicture}[font=\scriptsize]

\begin{scope}[yshift=4.8cm]
  \fill[blue!7] (0,0) rectangle (7.8,1.7);
  \foreach \y in {0.45,0.9,1.35}{
    \draw[blue!65!black,thick] (0.1,\y) -- (7.5,\y);
  }
  \draw[->,thick] (0,0)--(7.9,0) node[right]{$V$};
  \draw[->,thick] (0,0)--(0,1.85) node[above]{$\sigma$};
  \node[blue!70!black] at (6.5,1.5) {$S_2$};
  \node[blue!70!black] at (6.5,1.0) {$S_1$};
  \node[anchor=west] at (0.15,1.55) {\textbf{(a)} Classical:\;
    $T_i = \mathrm{const}$,\quad $g_{ij}=0$};
\end{scope}

\begin{scope}[yshift=2.55cm]
  \fill[blue!7]  (0,0) rectangle (7.8,1.7);
  \fill[blue!20,opacity=0.5] (4.5,0) rectangle (7.8,1.7);
  \fill[blue!35,opacity=0.4] (6.2,0) rectangle (7.8,1.7);
  \draw[blue!65!black,thick]
    (0.1,0.35) .. controls (2.5,0.45) and (5.0,0.70) .. (7.5,1.25);
  \draw[blue!65!black,thick]
    (0.1,0.85) .. controls (2.5,0.95) and (5.0,1.22) .. (7.5,1.62);
  \draw[blue!40!black,thick]
    (0.1,0.60) .. controls (2.5,0.68) and (5.0,0.95) .. (7.5,1.45);
  \draw[->,thick] (0,0)--(7.9,0) node[right]{$V$};
  \draw[->,thick] (0,0)--(0,1.85) node[above]{$\sigma$};
  \draw[gray!60,dashed,thick]
    (3.0,0.65)--(4.1,0.75)--(3.9,1.20)--(2.8,1.10)--cycle;
  \draw[->,red!80!black,line width=0.9pt]
    (3.45,0.95) arc[start angle=80,end angle=-190,radius=0.22];
  \node[red!80!black] at (4.7,1.38) {$\oint\omega_i\!\neq\!0$};
  \draw[->,orange!85!black,very thick]
    (5.8,0.55)--(6.55,1.05) node[midway,above,sloped]{$\nabla\!\zeta_V$};
  \node[anchor=west] at (0.15,1.55) {\textbf{(b)} MCTE:\;
    $\zeta_i T_i = C$,\quad $g_{ij}\neq 0$};
\end{scope}

\begin{scope}[yshift=0cm]
  \fill[blue!7]  (0,0) rectangle (7.8,1.7);
  \fill[blue!20,opacity=0.5] (5.5,0) rectangle (7.8,1.7);
  \fill[blue!38,opacity=0.38] (6.8,0) rectangle (7.8,1.7);
  \draw[->,thick] (0,0)--(7.9,0) node[right]{$\phi$};
  \draw[->,thick] (0,0)--(0,1.85) node[above]{$\zeta_V$};
  \draw[gray!55,dashed] (0,0.45)--(7.5,0.45)
      node[right,gray!70]{$1$};
  \draw[orange!85!black,very thick,smooth]
    (0.2,0.45)..controls(2.5,0.46)and(4.5,0.50)..(5.8,0.60)
              ..controls(6.4,0.72)and(6.9,1.05)..(7.3,1.62);
  \draw[gray!50,dashed] (7.3,0)
      node[below]{$\phi_{\!J}$} -- (7.3,1.62);
  \filldraw[orange!85!black] (7.3,1.62) circle(2pt);
  \draw[<->,black!60] (7.6,0.45)--(7.6,1.62)
      node[midway,right]{$O(1)$};
  \node[anchor=west] at (0.15,1.55)
      {\textbf{(c)} $\zeta_V(\phi)$: $O(1)$ correction at jamming};
\end{scope}

\end{tikzpicture}
\caption{Three-panel schematic of the MCTE framework (column width).
  \textbf{(a)}~Classical ($g_{ij}=0$): entropy level sets are straight lines; each intensity $T_i$ is independently uniform. \textbf{(b)}~MCTE ($g_{ij}\neq 0$): level sets curve; a closed loop accumulates a non-trivial Ruppeiner holonomy $\oint\omega_i\neq 0$ (red), and the coupling coefficient $\zeta_V$ grows in the direction of increasing off-diagonal curvature (orange).  The unique first integral of this flow is $\zeta_i T_i=C$. \textbf{(c)}~$\zeta_V$ as a function of packing fraction $\phi$: unity in the dilute limit, $O(1)$ departure at jamming $\phi_J$, explaining the observed rise in Rowe's $K_\mu$.}
\label{fig:entropymanifold}
\end{figure}

\textit{Granular realisation.}---In Edwards' statistical mechanics~\cite{Powders}, the role of temperature is played by compactivity $\chi$ ($\chi^{-1}=\partial S/\partial V$).  The Blumenfeld--Edwards extension~\cite{Blumenfeld} introduces angoricity $\boldsymbol{\mathcal{A}}$ ($\mathcal{A}_{ij}^{-1}=\partial S/\partial\sigma_{ij}$) conjugate to the stress $\boldsymbol{\sigma}$, yielding the two-channel ($N=2$) entropy surface $S(V,\boldsymbol{\sigma})$ with block Ruppeiner metric
\begin{equation}
  g = -\!\begin{pmatrix}
    \partial^2 S/\partial V^2 &
    \partial^2 S/\partial V\partial\sigma_{ij} \\[2pt]
    \partial^2 S/\partial\sigma_{ij}\partial V &
    \partial^2 S/\partial\sigma_{ij}\partial\sigma_{kl}
  \end{pmatrix}.
  \label{eq:granular-metric}
\end{equation}
\begin{remark}[System Parity]
Although this paper addresses an \emph{even-parity} system for illustration, where the skew-symmetric Onsager coupling block has no zero eigenvalue and inter-channel exchange is confined to closed reversible orbits, the general formalism is parity-agnostic.  The $N=2$ volume--stress granular ensemble is the canonical realisation;  subtleties specific to odd-parity ($N=3,5,\ldots$) systems, in which a topological zero mode opens an irreversibility mechanisms and drives non-associated plastic flow, will be treated in upcoming works.  Crucially, the $N=2$ MCTE equilibrium constitutes the thermodynamic \emph{background state} upon which the $N=3$ instability grows: shear-band nucleation requires a $N=3$ realisation, but its energetics are set by the $N=2$ Stable Layer from which it departs.
\end{remark}
The off-diagonal block $g_{V\sigma_{ij}}=-\partial\chi^{-1}/\partial \sigma_{ij}$ measures how much the inverse compactivity shifts when stress is varied at fixed volume.  It vanishes in the dilute limit and grows large near the jamming transition.  A sign subtlety merits emphasis: because $g_{ij} = -\partial^2 S/\partial q^i\partial q^j$, the off-diagonal entry $g_{V\sigma}$ is \emph{negative} when the cross-derivative $\partial^2 S/\partial V\partial\sigma$ is positive (stress-softening coupling).  Computing $g_{V\sigma}$ with the wrong sign breaks the MCTE invariance $\zeta_V\chi = C$; in the toy model of the Supplemental Material~\cite{supplement}, this error causes $\zeta_1 T_1$ to drift by $O(1)$ along the level set even though $S$ itself remains constant to machine precision, providing a stringent self-consistency diagnostic (see Supplemental Material, Section~S10). 
The correct equilibrium condition is $\zeta_V\chi=C$, \emph{not} $\chi=\mathrm{const}$.

\begin{figure}[tb]
  \centering
  \includegraphics[width=\columnwidth]{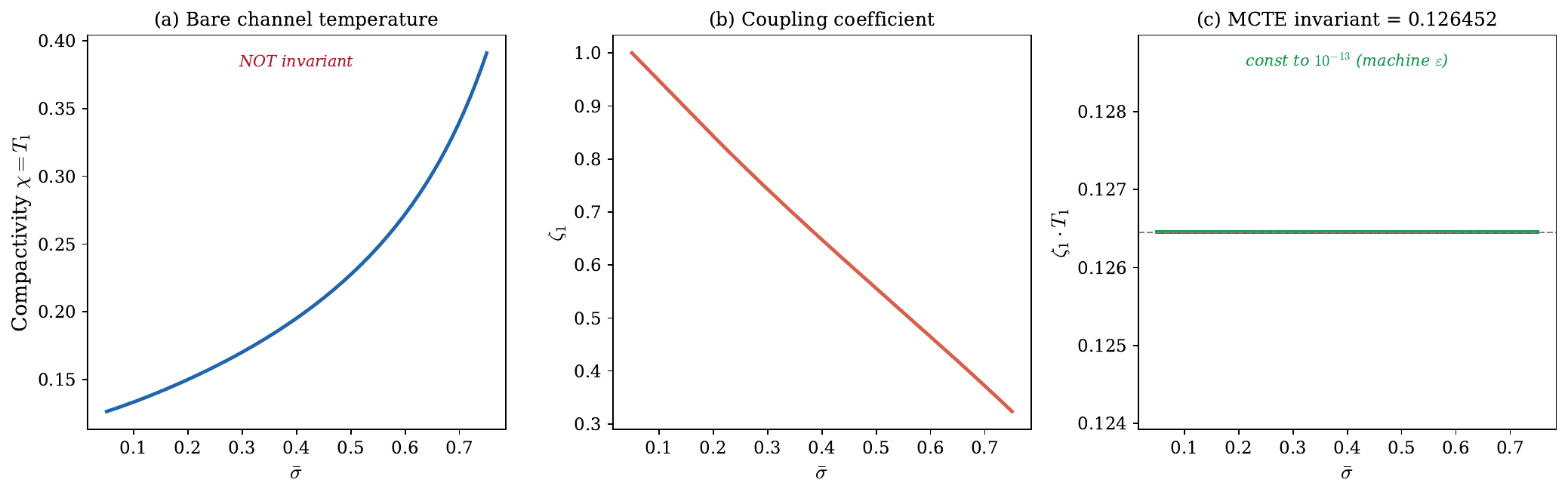}
  \caption{MCTE invariance on the toy entropy surface
  $S = a\ln(V\!-\!V_J) + b\ln(\sigma_{\max}\!-\!\sigma) - c\sigma/(V\!-\!V_J)$
  with $c=0.3$, $V_0=0.78$.
  (a)~Bare compactivity $\chi$ varies by a factor of~3 along the
  entropy level set.
  (b)~Coupling coefficient $\zeta_1$ compensates.
  (c)~Product $\zeta_1 T_1$ is constant to $10^{-13}$
  (machine precision), verifying the MCTE relation.
  Full parameter study in the Supplemental Material~\cite{supplement}.}
  \label{fig:invariant}
\end{figure}

Applying Eq.~(\ref{eq:zeta-explicit}) to the $(V,\sigma)$ system and projecting onto the normal direction $\hat{n}$ gives the scalar coupling ODE
\begin{equation}
  d\log\zeta_V =
  \frac{g_{VV}\mathcal{A}^{-1} - g_{V\sigma}\chi^{-1}}{\chi^{-2}}
  \,d\sigma,
  \label{eq:granular-zeta-ode}
\end{equation}
where $g_{V\sigma}\equiv\hat{n}_i\hat{n}_j g_{V\sigma_{ij}}$ and $\mathcal{A}\equiv\hat{n}_i\hat{n}_j\mathcal{A}_{ij}$ are the normal projections of the off-diagonal curvature and angoricity respectively. The fractional correction to the single-channel condition is
\begin{equation}
  \frac{|\nabla\zeta_V|}{\zeta_V}
  \sim \left|\frac{g_{V\sigma}}{g_{VV}}\cdot\frac{\chi}{\mathcal{A}}\right|,
  \label{eq:correction-estimate}
\end{equation}
with $g_{V\sigma}$ and $\mathcal{A}$ the normal projections defined above.
This correction is doubly suppressed in the dilute limit ($g_{V\sigma}\to 0$, $\chi/\mathcal{A}\to 0$ simultaneously) and becomes $O(1)$ near isostaticity: precisely where the coupled Blumenfeld--Edwards description is most needed, imposing $\chi$ and $\mathcal{A}$ independently introduces an $O(1)$ systematic error. Equation~(\ref{eq:correction-estimate}) is the first quantitative geometric estimate of this failure.  A concrete evaluation on a toy entropy surface $S(V,\sigma) = a\ln(V-V_J) + b\ln(\sigma_{\max}-\sigma) - c\sigma/(V-V_J)$, detailed in the Supplemental Material~\cite{supplement} (Sections~S8--S13), confirms that $\zeta_1 T_1 = \mathrm{const}$ holds to machine precision ($\sim 10^{-13}$) and yields the following quantitative benchmarks: for moderate coupling ($c=0.3$), the total MCTE correction $|\zeta_1-1|$ reaches 53\% at $V_0=0.80$ and 63\% near jamming ($V_0 - V_J = 0.04$); the purely cross-coupling contribution $|\zeta_1(c) - \zeta_1(0)|$ (the irreducible multi-channel correction absent in single-channel theory) reaches 17\% near jamming, rising to 28\% at $c=0.6$. 
Figure~\ref{fig:invariant} illustrates the key result.

\textit{Microscopic origin of the coupling.}---The growth of $g_{V\sigma}$ near jamming has a concrete dynamical origin. Biktashev and Tsyganov~\cite{BiktashevTsyganov2016} showed that adiabatic elimination of a fast diffusing intermediate species $w$ from a three-component self-diffusion system generates effective cross-diffusion $D_{uv},D_{vu}\neq 0$ in the reduced two-component system (their Eq.~12), with magnitude proportional to $D'_w=D_w(1-k^2D_w)$.  Via the fluctuation-dissipation relation, off-diagonal diffusivity is the kinetic dual of off-diagonal Ruppeiner curvature: $g_{V\sigma}\neq 0$ arises precisely because the fast intermediate channel, i.e. the mechanical fabric degree of freedom, $M$, has been adiabatically eliminated from the full $N=3$ system, imprinting its coupling onto the $N=2$ entropy surface.  The mapping is structural: both systems share the same three-channel topology and the same elimination mechanism, though the physical variables differ.

Two testable consequences follow directly.  First, the MCTE correction~(\ref{eq:correction-estimate}) becomes $O(1)$ at the diffusion length $\ell\sim D_w^{1/2}$ of the fast fabric channel, providing a microscopic length scale for the onset of the $\zeta_V K_\mu$ departure in Prediction~3 below.  Second, the B-T system supports quasi-soliton waves precisely at the wavenumber $k\sim D_w^{-1/2}$ where $D'_w\to 0$~\cite{BiktashevTsyganov2016}; in the granular context this predicts that spatially structured precursors to shear-band formation appear at the same length scale $\ell$ as the MCTE correction, a coincidence testable in DEM simulations.  The even-parity ($N=2$) framework here describes the equilibrium structure of the reduced system; the $N=3$ dynamics of the full system, which drives non-associated flow, will be addressed in detail in upcoming works.

\textit{Three predictions for granular plasticity.}---All three results below follow from the $N=2$ MCTE geometry without additional constitutive input; full proofs are in the Supplemental Material~\cite{supplement} (Sections~S1--S6).

\textit{Prediction 1 (dilatancy ratio).}  Along a quasi-static path maintaining $\zeta_V\chi=C$ on $S(V,\boldsymbol{\sigma})$, the dilatancy ratio in the normal direction $\hat{n}$ is
\begin{equation}
  D = -\frac{dV/V}{d\sigma/\sigma}
    = \frac{g_{V\sigma}}{g_{VV}}\cdot\frac{\chi}{\mathcal{A}}
    + O\!\left(\!\left(\tfrac{g_{V\sigma}}{g_{VV}}\right)^{\!2}\right),
  \label{eq:dil-ruppeiner}
\end{equation}
where $g_{V\sigma}\equiv\hat{n}_i\hat{n}_jg_{V\sigma_{ij}}$. Dilatancy \emph{is} the off-diagonal Ruppeiner curvature, weighted by the ratio of intensive variables.

\textit{Prediction 2 (Rowe's energy restriction).}  The non-negativity of frictional dissipation in Rowe's theorem~\cite{Rowe1962}, $0\leq D_R\leq 1-K_\mu^{-1}$, is equivalent to positive semi-definiteness of the $2\times 2$ Ruppeiner metric:
\begin{equation}
  \det\begin{pmatrix}g_{VV}&g_{V\sigma}\\g_{\sigma V}&g_{\sigma\sigma}
  \end{pmatrix}\geq 0.
  \label{eq:Ruppeiner-stability}
\end{equation}
Equality is the critical state ($\det g=0$, maximal channel coupling, maximum $|\zeta_V-1|$).  Rowe's restriction is therefore not an independent constitutive hypothesis but a consequence of thermodynamic stability ($\delta^2 S\leq 0$): any admissible entropy surface automatically satisfies the Rowe bound.

\textit{Prediction 3 (MCTE-corrected Rowe relation).}  Rowe's stress-dilatancy law~\cite{Rowe1962}, $\sigma_1'/\sigma_3'=K_\mu R$, has been observed for 60 years to require a state-dependent $K_\mu$ near jamming, with no geometric explanation~\cite{Blumenfeld}.  The MCTE framework resolves this directly: the correct relation is
\begin{equation}
  \frac{\sigma_1'}{\sigma_3'} = \zeta_V(\boldsymbol{\sigma})\,K_\mu\,R,
  \label{eq:MCTE-Rowe}
\end{equation}
where $\zeta_V(\boldsymbol{\sigma})$ is computable from $S(V,\boldsymbol{\sigma})$ alone with no free parameters.  In the decoupled limit $\zeta_V\to 1$, recovering the classical result exactly.  Near jamming $|\zeta_V-1|=O(g_{V\sigma}/g_{VV})=O(1)$, and $\zeta_V K_\mu$ departs from the contact-scale constant by an amount set entirely by the off-diagonal entropy curvature.  At the critical state, $\zeta_V$ achieves its maximum consistent with $\det g=0$, and Eq.~(\ref{eq:MCTE-Rowe}) predicts the critical-state stress ratio $M$ from geometry alone.

The three predictions are testable without free parameters: extract $S(V,\boldsymbol{\sigma})$ from discrete-element simulations of the Blumenfeld--Edwards ensemble, compute $g_{ij}$ from Eq.~(\ref{eq:granular-metric}), and compare $D$, the Rowe bound, and $\zeta_V K_\mu$ directly with simulation data.  Crucially, $\zeta_V(\boldsymbol{\sigma})$ in Eq.~(\ref{eq:MCTE-Rowe}) is extracted from the \emph{local} curvature of the entropy surface at each stress state, making the prediction spatially resolved: different regions of a heterogeneous packing will have different $\zeta_V$ values, and the spatial map of $\zeta_V K_\mu$ is a direct observable distinguishing the MCTE correction from any global (spatially uniform) empirical rescaling of $K_\mu$.  The $N=2$ framework has no zero mode, so these results hold independently of questions about non-associated flow, which require the odd-channel ($N=3$) extensions.

\textit{Conclusions.}---We have shown that when multiple thermodynamic channels are non-negligibly coupled, the correct equilibrium condition is $\zeta_i T_i=C$: a single conserved scalar, with coupling coefficient $\zeta_i$ equal to the holonomy of the Ruppeiner connection on the entropy level set.  This generalises both the classical uniform-temperature condition and the relativistic Tolman temperature, and reduces to each in the appropriate limit.  For the two-channel granular ensemble, three classical results of granular plasticity: Rowe's dilatancy ratio, his energy restriction, and the state-dependence of $K_\mu$ emerge as parameter-free geometric consequences of the off-diagonal Ruppeiner curvature.

There are two most direct experimental tests: The \emph{$C$-uniformity test}: across a developing shear band the individual intensities $\chi$, $\mathcal{A}$ should be spatially non-uniform, while $\zeta_V\chi$ remains uniform, accessible to continuum simulations and discriminating the MCTE picture from any classical multi-temperature alternative.  The \emph{length-scale test}: the onset of the $O(1)$ departure in $\zeta_V K_\mu$ should occur at the diffusion length $\ell\sim D_w^{1/2}$ of the fast fabric channel, matching the quasi-soliton wavenumber in the dual reaction-diffusion system~\cite{BiktashevTsyganov2016} and providing a parameter-free prediction for DEM simulations.

\begin{acknowledgments}
The authors acknowledge support from the ARC Centre of Excellence for
Carbon Science \& Innovation (grant CE230100021) and Curtin University.
\end{acknowledgments}

\bibliographystyle{apsrev4-2}
\bibliography{apssamp}

\end{document}